\documentclass{ws-procs9x6}
\begin{document}

\title{Formation of small-scale structure in SUSY CDM}

\author{S.~HOFMANN$^\sharp$, D.~J.~SCHWARZ$^\flat$ and H.~ST\"OCKER$^\sharp$}

\address{
${}^\sharp$Institut f\"ur Theoretische Physik, 
Johann Wolfgang Goethe-Universit\"at, \\
60054 Frankfurt am Main, Germany\\
${}^\flat$Theory Division, CERN, 1211 Geneva 23, Switzerland \\
E-mail: stehof@th.physik.uni-frankfurt.de, dominik.schwarz@cern.ch,
stoecker@th.physik.uni-frankfurt.de}

\maketitle

\abstracts{
The lightest supersymmetric particle, most likely the lightest neutralino,
is one of the most prominent particle candidates for cold dark matter (CDM).
We show that the primordial spectrum of density fluctuations in neutralino 
CDM has a sharp cut-off, induced by two different damping mechanisms. 
During the kinetic decoupling of neutralinos, non-equilibrium processes  
constitute viscosity effects, which damp or even absorb density perturbations
in CDM. After the last scattering of neutralinos, free streaming induces 
neutralino flows from overdense to underdense regions of space.
Both damping mechanisms together define a minimal mass scale for 
perturbations in neutralino CDM, before the inhomogeneities enter the 
nonlinear epoch of structure formation.
We find that the very first gravitationally bound neutralino clouds ought to 
have masses above $10^{-6} M_\odot$, which is six orders of magnitude above the
mass of possible axion miniclusters. 
}

\section{Introduction}

Recent measurements support the idea that there is a significant amount of 
cold dark matter (CDM) in the Universe. An analysis of the temperature 
anisotropies in the cosmic microwave background (CMB) gives for the CDM 
mass density $\Omega_{\rm cdm} h^2 = 0.13^{+0.03}_{-0.02}$ (all data with 
weak $h$-prior)\cite{Sievers}, while from the same analysis the mass density 
of the baryons is much smaller $\Omega_{\rm b} h^2 = 0.023^{+0.003}_{-0.003}$.
The latter is consistent with the measurement of the primordial deuterium 
abundance in high-redshift hydrogen clouds, which gives the most precise 
determination of the baryon density\cite{OMeara}, 
$\Omega_{\rm b} h^2 = 0.0205 \pm 0.0018$. 
Large galaxy redshift surveys support the CMB 
measurements: from the analysis of 160,000 galaxies Percival et al\cite{per01}
find $(\Omega_{\rm b} + \Omega_{\rm cdm}) h = 0.20 \pm 0.03$
and $\Omega_{\rm b} /(\Omega_{\rm b} + \Omega_{\rm cdm}) = 0.15 \pm 0.07$.
Taking various observations together, Turner\cite{Turner} estimates that 
the non-baryonic mass in the Universe, which is just the mass of CDM,
is given by $\Omega_{\rm cdm} = 0.29\pm 0.04$.
 
The characteristic feature of CDM is its non-relativistic equation of state 
at the time when the Universe contains enough matter within one Hubble volume
to form one galaxy. The two leading particle candidates for CDM are the axion,
a very light particle with a mass in the range $m_a \sim (10^{-6}-10^{-5})$ eV,
which is cold since it has been produced in a Bose condensate, and the 
neutralino, a heavy particle with a mass $M_{\tilde{\chi}} \sim (50-500)$ GeV, 
which is cold by virtue of its large mass. In the constrained minimal 
supersymmetric extension of the standard model (MSSM), the lightest neutralino 
is most likely the bino.

For almost 20 years there has been a strong working hypothesis:
axion CDM and neutralino CDM cannot be distinguished by purely cosmological 
observations because both particle candidates interact only via gravity, as 
far as cosmology goes. This fact has made the study of structure formation 
on large scales $> 1$ Mpc simple---the microphysics of CDM is irrelevant.
However, at the galactic scale and below, various particle candidates might 
be distinguishable. To learn more about the nature of CDM,
we study the small-scale structure of neutralino CDM with emphasis on
the very first, purely gravitationally bound, neutralino clouds
(for details, see our recent work\cite{hof01}).

\section{Chemical and kinetic decoupling}

\begin{figure}[ht]
\centerline{\epsfxsize=2.5in\epsfbox{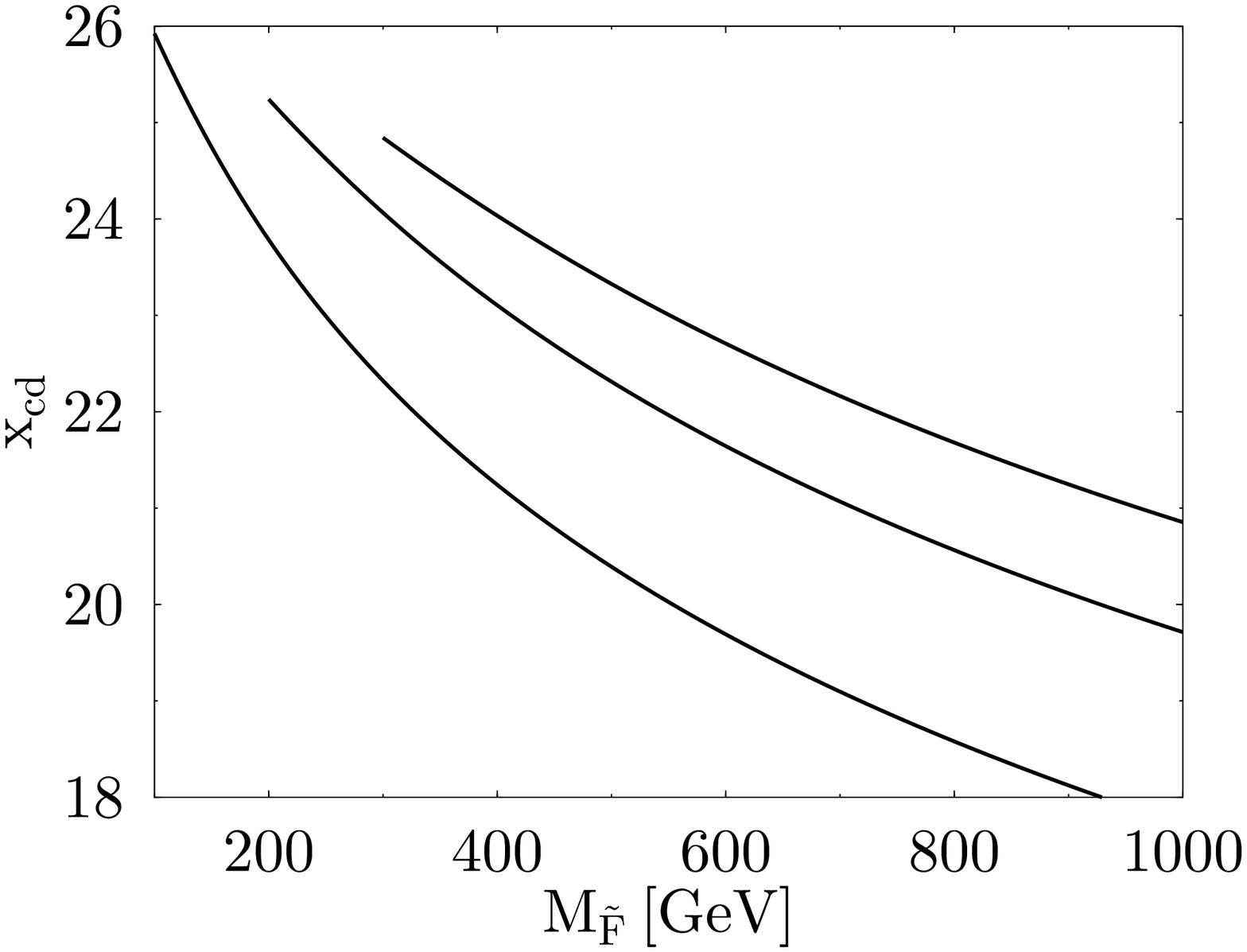}
\epsfxsize=2.5in\epsfbox{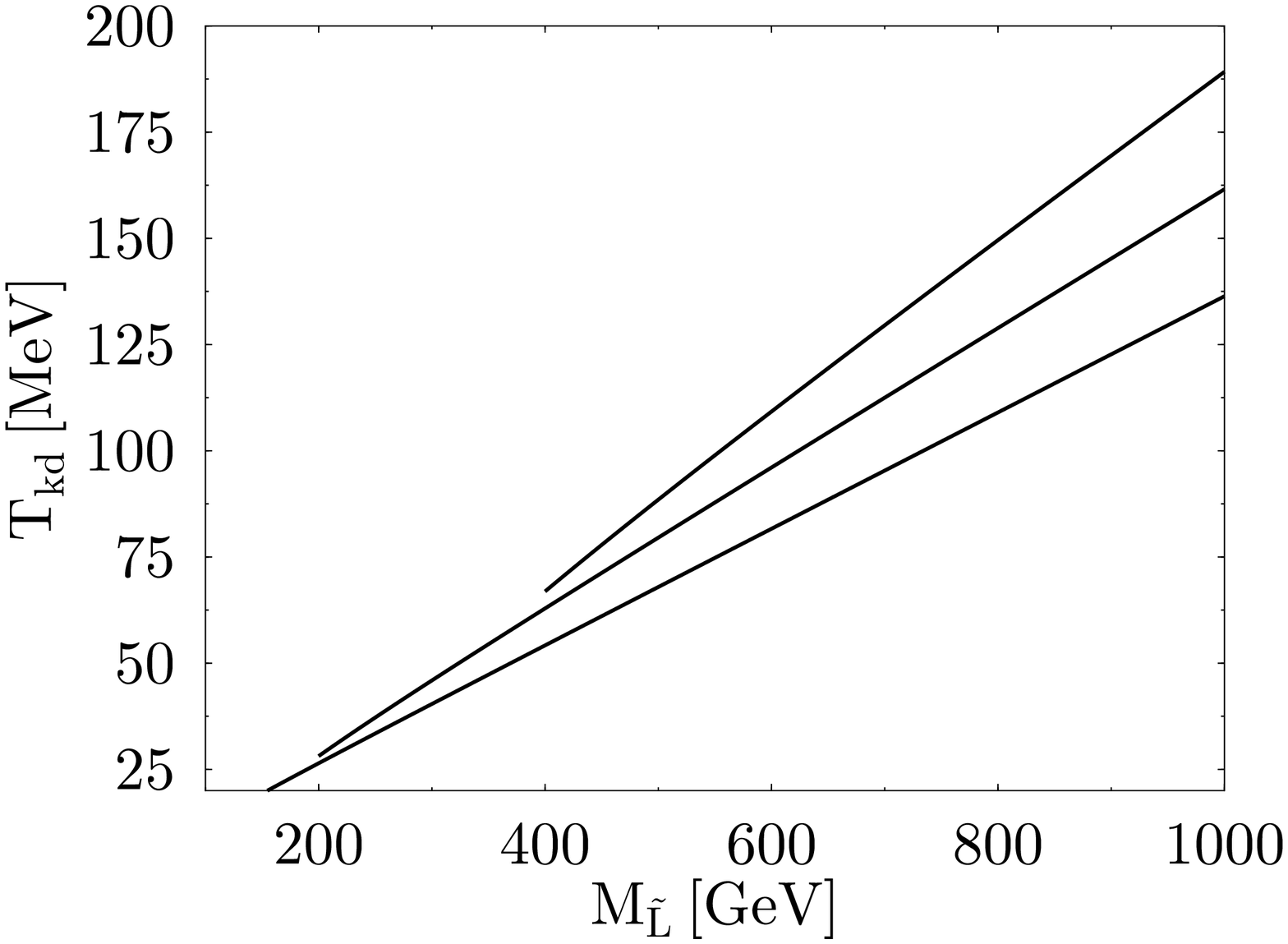}}   
\caption{Chemical and kinetic decoupling 
as a function of the universal sfermion (slepton) mass
for bino masses $M_{\widetilde{\chi}}\in\{50,100,200\}$ GeV
(from bottom to top).
\label{dec}}
\end{figure}

There are two distinct temperature scales for CDM particles that are massive 
and weakly interacting and obey a thermal history. For temperatures 
$T>T_{\rm cd}$, binos are kept in chemical equilibrium with all fermions in 
the heat bath via the annihilation processes 
$\widetilde{\chi} + \widetilde{\chi} \leftrightarrow \overline{F} + F$,
at the rate
\begin{equation}
\Gamma_{\rm ann} \approx 10^{-3} \; M_{\widetilde{\chi}} \;
\frac{M^{\;\; 4}_{\widetilde{\chi}}}{\left(M^{\; \; 2}_{\widetilde{F}}
+M^{\; \; 2}_{\widetilde{\chi}}\right)^2} \;
\frac{M^{\;\; 4}_{\widetilde{F}}+
M^{\;\; 4}_{\widetilde{\chi}}}{\left(M^{\;\;2}_{\widetilde{F}}+
M^{\;\; 2}_{\widetilde{\chi}}\right)^2}
\; x^{-\frac{5}{2}} \; {\rm e}^{-x} \; .
\end{equation}
Here, $x = M_{\widetilde{\chi}} / T$ and $M_{\widetilde{F}}$
denotes the universal sfermion mass.
The numerical prefactor is the effective neutralino--fermion coupling,
the second factor shows the dependence on the MSSM parameters
and the $x$-dependent factor gives the temperature dependence
in units of the neutralino mass and is proportional to
the bino number density.
Chemical decoupling of binos happens at the temperature 
$T_{\rm cd} = M_{\widetilde{\chi}} /x_{\rm cd}$,
when the neutralino annihilation rate becomes comparable to the
Hubble rate:
\begin{equation}
x_{\rm cd} 
\approx 
{\rm ln}
\left[ 10^{-4} \; \frac{M_{\rm Pl} \left(
M^{\;\; 4}_{\widetilde{F}}+M^{\;\; 4}_{\widetilde{\chi}}
\right) M_{\widetilde{\chi}}^{\;\; 3}}{
\left(M^{\;\; 2}_{\widetilde{F}}+M^{\;\; 2}_{\widetilde{\chi}}\right)^4}
\right]
\; .
\end{equation}
Scanning the MSSM parameter space, we typically find $x_{\rm cd} \approx 22$,
see Figure~\ref{dec}. The chemical decoupling temperature $T_{\rm cd}$ 
increases with increasing bino mass, since the neutralino number density
is suppressed with $\exp{(-M_{\widetilde{\chi}} /T)}$. As a consequence,
the annihilation rate approaches the Hubble rate faster for larger
bino masses. For a fixed bino mass, $T_{\rm cd}$ increases for
an increasing sfermion mass, since the interaction range decreases. 
(Note that this discussion is not correct on a quantitative level, since we
neglect here important effects like co-annihilation, in order to render an
analytic discussion possible.)

For temperatures $T_{\rm kd}<T<T_{\rm cd}$ neutralinos are in
local thermal equilibrium (lte) with all fermions of the heat bath.
Equilibrium is maintained due to elastic scattering processes
$\widetilde{\chi} + \{\overline{F},F\} \leftrightarrow 
\widetilde{\chi} + \{\overline{F},F\}$
at the rate
\begin{equation}
\Gamma_{\rm el}  \approx  10^{-2} \; M_{\widetilde{\chi}}  \;
\frac{M^{\;\; 4}_{\widetilde{\chi}}}
{\left(M^{\;\; 2}_{\widetilde{\rm F}}-M^2_{\widetilde{\chi}}\right)^2}  
\; x^{-5}
\; .
\end{equation}
Kinetic decoupling of binos takes place 
when the relaxation time $\tau_{\rm relax}$ in the bino system exceeds the 
Hubble time.
The relaxation time $\tau_{\rm relax} = N \tau_{\rm coll}$
differs from the collision time
$\tau_{\rm coll} = 1/\Gamma_{\rm el}$ by the number $N(T)$
of elastic scatterings needed to keep or establish lte 
in the bino system.
This number is given by the relative momentum transfer 
per elastic scattering, 
$N(T) = (\Delta P_{\widetilde{\chi}} / P_{\widetilde{\chi}})^{-1}
\approx M_{\widetilde{\chi}} / T$, with
$\Delta P_{\widetilde{\chi}} = \sqrt{<t>} \approx 3 T/\sqrt{2}$
denoting the rms of the Mandelstam variable $t$. 
Kinetic decoupling happens at
\begin{equation}
T_{\rm kd}
\approx
\left[
10^2 \; \frac{M_{\widetilde{\chi}}^{\;\; \alpha}
\left(M^{\;\; 2}_{\widetilde{F}} - M^{\;\; 2}_{\widetilde{\chi}}\right)^2}
{M_{\rm Pl}}\right]^\frac{1}{3+\alpha}
\; ,
\end{equation}
with $\alpha = 0$ if mistakenly $\tau_{\rm coll} \equiv \tau_{\rm relax}$
is assumed and $\alpha = 1$ when the number of scatterings $N$ is taken
into account. We typically find $T_{\rm kd} = (10-100)$ MeV, see
Figure~\ref{dec}. The kinetic decoupling temperature is increasing
with increasing bino mass because the momentum transfer 
to the heat bath is decreasing. As a consequence the number of elastic
scattering processes needed to keep or establish lte is
increasing.

We find $T_{\rm cd} \gg T_{\rm kd}$, because
$\Gamma_{\rm ann} / \Gamma_{\rm el} \approx 10^{-1} x^{5/2} \exp{(-x)} \ll 1$
for $T < T_{\rm cd} < M_{\widetilde{\chi}}$; the number density of 
neutralinos is Boltzmann suppressed with respect to the number densities 
of the relativistic fermions. This is the reason for the temperature hierarchy 
$T_{\rm cd} \gg T_{\rm kd} > T_{\rm ls}$, with $T_{\rm ls}$ denoting the 
temperature at which binos scatter for the last time.

\section{Local transport coefficients}

During the process of kinetic decoupling, non-equilibrium processes
constitute themselves as viscosity phenomena in the bino system. 
We therefore use hydrodynamics for the description.
For $T\gg T_{\rm cd}$, CDM and the heat bath can be described by
a single ideal fluid. For $T_{\rm cd} > T > T_{\rm kd}$ 
the CDM fluid is strongly coupled to the radiation fluid (rad),
which keeps CDM in lte. Around $T_{\rm kd}$, the CDM fluid
decouples from the radiation fluid, in which lte persists,
since $\Omega_{\rm cdm} = (a/a_{\rm eq}) \Omega_{\rm rad} \ll 
\Omega_{\rm rad}$ for $T\gg T_{\rm eq}$.

The resulting non-equilibrium processes in the CDM fluid
can be taken into account by additional Lorentz tensors
${\bf J}^{(1)}$ and ${\bf T}^{(1)}$
in the current density ${\bf J}_{\rm cdm}$ and the energy momentum
tensor ${\bf T}_{\rm cdm}$ of the CDM fluid.
We fix the ambiguities in the relativistic description of imperfect fluids
by demanding ${\bf J}^{(1)} \equiv {\bf 0}$ and
${\bf T}^{(1)} = \zeta \; {\bf h} {\bf \nabla}\cdot {\bf U} 
+\eta \; {\bf W}^{({\rm T})}+
\chi \; {\rm Sym}\left({\bf U}\otimes {\bf Q}^{({\rm T})}\right)$.
${\bf U}$ denotes the adiabatic velocity field and ${\bf h}$
projects on the hypersurface perpendicular to it.
The first term is the bulk viscosity, describing the flow
of ${\bf U}$ in the hypersurface defined by ${\bf h}$.
The second term is the shear viscosity and describes the bending
of ${\bf U}$ in the direction perpendicular to the adiabatic current.
The last term is the heat conduction.

The strength of the dissipative processes is given by the
local transport coefficients $\zeta$, $\eta$ and $\chi$.
An efficient method for calculating these coefficients was proposed
by Silk\cite{sil68}. It has been applied in great detail to the case of
a relativistic fluid by Weinberg\cite{wei71}. 
A generalisation to an arbitrary equation of state can be found in 
our recent work\cite{hof01}. 

The main idea is to compare the Lorentz tensors 
in the hydrodynamical description with the Lorentz tensors in
the kinetic description. In the kinetic description we make the
ansatz 
$F^{(1)}(\omega,n,x) = {\bf A}(\omega,x) + {\bf B}(\omega,x) \cdot {\bf n}
+
{\bf C}(\omega,x) \cdot \left({\bf n}\otimes {\bf n} + 1/3\, {\bf h}\right)$
for the CDM phase-space distribution describing 
the non-equilibrium state of the bino system.
Here, $\omega$ denotes the projection of the momentum on the
velocity field and ${\bf n}$ denotes a vector perpendicular
to the adiabatic current. We calculate\cite{hof01} the coefficients in 
the expansion of $F^{(1)}$ into irreducible polynoms in ${\bf n}$ and
${\bf h}$. For the tensor structure we find
${\bf A} \propto {\bf \nabla}\cdot {\bf V}$, 
${\bf B} \propto {\bf Q}^{({\rm T})}$ and 
${\bf C} \propto {\bf W}^{({\rm T})}$.
The scalar ${\bf A}$ generates dissipative processes, which are not
perpendicular to the adiabatic current. This contribution deserves
special care.

Calculating the energy momentum tensor in the kinetic description
and comparing it with ${\bf T}^{(1)}$, we find the local
transport coefficients $\zeta = 5\rho_{\rm cdm} /3 \Gamma_{\rm el}$,
$\eta = \rho_{\rm cdm} /\Gamma_{\rm el}$ and $\chi \equiv 0$
in first order in $1/\Gamma_{\rm el}$ and $1/x$.
All coefficients are decreasing with an increasing elastic
scattering rate, since elastic scattering allows the transfer
of derivations from lte to the heat bath. Heat conduction
is a subdominant process for non-relativistic particles
and vanishes in leading order.

\section{Acoustic absorption}

\begin{figure}[ht]
\centerline{\epsfxsize=2.5in\epsfbox{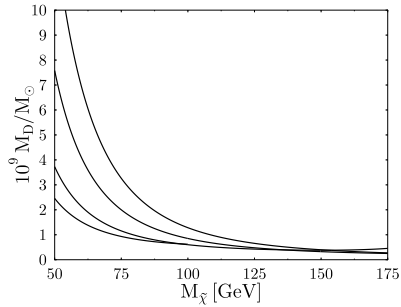}
}
\caption{Acoustic damping scale as a function of the bino mass
for sfermion masses $M_{\widetilde{F}}\in\{150,200,300,400\}$ GeV
(from bottom to top).
\label{abs}}
\end{figure}

The dissipative processes presented in the last section
transfer energy and momentum in the direction perpendicular
to the adiabatic flow. This provides a damping mechanism
for acoustic perturbations in CDM\cite{wei71}. 
The damping of density inhomogeneities is given by 
${\rm Im}(\omega)$ as calculated in linear perturbation theory
for relativistic hydrodynamics.
For an acoustic wave with wave number $k$ we find
\begin{equation}
\frac{\delta \rho_{\widetilde{\chi}}}{\rho_{\widetilde{\chi}}}
\propto
\exp{\left[
-\frac{3}{2} \int\limits_0^{t(T_{\rm kd})} {\rm d}t
\; \frac{k^2}{\Gamma_{\rm el}}\right]}
=
\exp{\left[\left(
-\frac{M_{\rm d}}{M}\right)^{2/3}\right]}
\; ,
\end{equation}
where  
$M_{\rm d} \approx 3\cdot 10^{-8} 
({\rm GeV}^2/M_{\widetilde{\chi}} T_{\rm kd})^{3/2} 
(\Omega_{\widetilde{\chi}} h^2) M_\odot$
is the characteristic damping scale.
We find $M_{\rm d} \approx 10^{-9} M_\odot$, see Figure~\ref{abs}.
Thus, only acoustic perturbations with masses $M>M_{\rm d}$, contained 
in the overdense volume, are not absorbed and enter the free streaming regime.

\section{Free streaming}

For temperatures $T < T_{\rm ls}$, CDM does freely streaming on geodesics.
As a consequence, CDM propagates
from overdense to underdense
regions, thus smearing out local inhomogeneities.
We find for the induced damping scale 
$M_{\rm fs}(a) = M_{\rm d} {\rm ln}^3(a/a_{\rm ls})$.
Free streaming becomes the leading damping mechanism
once the universe has doubled its size after last scattering.
It is interesting to calculate the free streaming scale
at the time of matter-radiation equality, when CDM density perturbations
start to grow linearly with the expansion factor.
We typically find $M_{\rm fs}(a_{\rm eq}) \approx 10^{-6} M_\odot$.

\section{Conclusions}

We have shown that collisional damping 
and free streaming smear out all power of primordial
density inhomogeneities in bino CDM below $10^{-6} M_\odot$ by the
time of matter-radiation equality.
This is in striking contrast to claims in the literature\cite{gur97}
that the minimal mass for the first purely gravitationally bound
neutralino cloud is 
$10^{-13} \, (150 {\rm\ GeV}/M_{\widetilde{\chi}})^3 \; M_\odot$.
The huge difference to our result stems from the assumption that
chemical and kinetic decoupling happened simultaneously, which
we proved to be wrong. We find instead that the very first bino objects
have to have masses above $10^{-6} M_\odot$. This result is very robust 
with respect to the MSSM parameters. 
These bino clouds are very different from possible axion 
mini-clusters with typical masses\cite{kol96} around $10^{-12} M_\odot$.  

According to hierarchical structure formation these very first
CDM clouds are supposed to merge and form larger objects.
Large scale structure simulations show structure formation
on all accessible scales down to the resolution of the 
simulation\cite{kamp00}.
However, the dynamic range of todays simulations is not sufficient to deal 
with the very first CDM objects, so the fate of these objects is an open issue.
A cloudy distribution of CDM in galactic halos would have important
implications for direct\cite{anne02}
and indirect\cite{berg} searches for dark matter.

\section*{Acknowledgements}
S.~H. thanks J.~Edsj\"o, G.~Fuller, A.~Greene, P.~Ullio and J.~Silk
for valuable comments and suggestions and acknowledges financial
support of ``Vereinigung von Freunden und F\"orderern
der Johann Wolfgang Goethe-Universit\"at Frankfurt am Main e.V.''
and the Marie Curie fellowship.

\end{document}